\documentclass[conference]{IEEEtran}
\IEEEoverridecommandlockouts
\pdfoutput=1
\usepackage{cite}
\usepackage{amsmath,amssymb,amsfonts}
\usepackage{algorithmic}
\usepackage{graphicx}
\usepackage{textcomp}
\usepackage{xcolor}
\usepackage{multirow}
\usepackage{threeparttable}
\usepackage{booktabs}
\usepackage{array}
\usepackage[hidelinks]{hyperref}
\usepackage[caption=false,font=footnotesize]{subfig}
\def\BibTeX{{\rm B\kern-.05em{\sc i\kern-.025em b}\kern-.08em
    T\kern-.1667em\lower.7ex\hbox{E}\kern-.125emX}}
\begin{document}

\title{Efficient Stochastic Polar Decoder \\ With Correlated Stochastic Computing\\
}

\author{\IEEEauthorblockN{Jiaxing Li}
\IEEEauthorblockA{\textit{School of Artificial Intelligence} \\
\textit{Beijing University of}\\\textit{ Posts and Telecommunications}\\
BeiJing, China \\
lijiaxing@bupt.edu.cn}
\and
\IEEEauthorblockN{Zhisong Bie\textsuperscript{*}}
\IEEEauthorblockA{\textit{School of Artificial Intelligence} \\
\textit{Beijing University of}\\\textit{ Posts and Telecommunications}\\
BeiJing, China \\
zhisongbie@bupt.edu.cn}
\and
\IEEEauthorblockN{Shuwen Zhang}
\IEEEauthorblockA{\textit{School of Artificial Intelligence} \\
\textit{Beijing University of}\\\textit{ Posts and Telecommunications}\\
BeiJing, China \\
zhangshuwen@bupt.edu.cn}
}

\maketitle

\begin{abstract}
Polar codes have gained significant attention in channel coding for their ability to approach the capacity of binary input discrete memoryless channels (B-DMCs), thanks to their reliability and efficiency in transmission. However, existing decoders often struggle to balance hardware area and performance. Stochastic computing offers a way to simplify circuits, and previous work has implemented decoding using this approach. A common issue with these methods is performance degradation caused by the introduction of correlation. This paper presents an Efficient Correlated Stochastic Polar Decoder (ECS-PD) that fundamentally addresses the issue of the `hold-state', preventing it from increasing as correlation computation progresses. We propose two optimization strategies aimed at reducing iteration latency, increasing throughput, and simplifying the circuit to improve hardware efficiency. The optimization can reduce the number of iterations by 25.2\% at $\boldsymbol{E_b/N_0}$ = 3 dB. Compared to other efficient designs, the proposed ECS-PD achieves higher throughput and is 2.7 times more hardware-efficient than the min-sum decoder.
\end{abstract}

\renewcommand{\thefootnote}{\fnsymbol{footnote}}
\footnotetext[1]{Corresponding author}


\section{Introduction}
\label{sec:1}
Polar codes, initially proposed by Arikan\cite{Arikan2009}, have become popular in channel coding due to their ability to achieve the capacity of binary-input discrete memoryless channels (B-DMCs) . Furthermore, polar codes have been selected for channel coding in the control channels of the Third Generation Partnership Project (3GPP) New Radio (NR)\cite{Hui2018}, owing to their reliability and efficiency of transmission. The primary decoding methods for polar codes include Successive Cancellation (SC) and Belief Propagation (BP) decoding. SC decoding is notable for its low complexity of $\mathcal{O}(N\log {N})$, but it operates serially, which can limit throughput. In response, Successive Cancellation List (SCL) decoding was proposed to enhance error-correction performance, albeit with the trade-off of higher complexity, specifically $\mathcal{O}(LN\log {N})$\cite{Tal2015}. BP decoding, on the other hand, operates in parallel, offering higher throughput, which is advantageous for applications requiring fast decoding speeds. The complexity of BP decoding is $\mathcal{O}(IN\log {N})$, where $I$ is the number of iterations, necessitating strategies to reduce this complexity. Thus, a balance between decoding performance and implementation complexity remains a challenge.

Some strategies have been suggested to improve the efficiency of polar decoders, such as the min-sum algorithm\cite{Abbas2017}, bit flipping\cite{Zhao2023}, early stopping\cite{Simsek2016}, and stage-combined BP\cite{Hasani2022}. These techniques strive to balance performance and complexity. However, most of these methods emphasize deterministic computation.
Recently, stochastic computing has garnered significant interest due to its benefits in error tolerance and hardware efficiency compared to deterministic computing. Stochastic computing uses random bit streams to represent numbers, simplifying complex operations. These benefits align well with BP polar code decoders, significantly reducing resource usage in parallel architectures. Stochastic polar code decoders were initially introduced using the SC algorithm and presented stochastic computing components\cite{Xu2014}. However, these decoders suffered significant performance loss compared to deterministic computation. Subsequent work proposed stochastic BP decoders\cite{Yuan2016}\cite{Xu2016}, but correlation in bit streams during iterations, as determined by the probability update formula, presented challenges. Xu addressed this issue by developing a rerandomization module and adopting an early termination technique to enhance convergence speed and reduce decoding delay\cite{Xu2020}.


Despite these advancements, high Signal-to-Noise Ratio (SNR) conditions still lead to a significant number of `hold-state', limiting the minimum error rate \cite{Sharifi2008}. This paper presents an Efficient Correlated Stochastic Polar Decoder (ECS-PD) to mitigate this challenge. 
The principal contributions of this work can be summarized as follows:
\begin{itemize}
\item[$\bullet$] The proposed decoder utilizes correlated stochastic computing to address the “hold-state” issue inherent in traditional stochastic decoding. By operating in the log-likelihood ratio (LLR) domain, the design achieves improved error tolerance and lower complexity, making it highly efficient for hardware implementation.
\item[$\bullet$] 
This paper introduces simplified computing units (CUs) to optimize the decoding process. These simplified CUs reduce complexity by efficiently handling frozen bits and minimizing unnecessary computations, leading to lower latency and higher throughput.
\item[$\bullet$] 
The hardware design of ECS-PD introduces an optimized architecture based on stochastic computing, achieving a throughput of \( 3816.5 \, \text{Mb/s} \) at \( 1625 \, \text{MHz} \) and hardware efficiency of \( 25,614 \, \text{Mb/s/mm}^2 \) at $E_b/N_0=3.5\,\text{dB}$. This design significantly minimizes resource usage and area, making it highly efficient for practical implementations.
\end{itemize}


\section{Polar Decoding Algorithm And Stochastic Computing}
\label{sec:2}
\subsection{Polar Codes}
The polar code $\textbf{u}_{1}^{N}\! \in \mathbb{C}^{n \times 1}=\{u_{1}, \ldots, u_{n}\}$ is encoded into $\textbf{x}_{1}^{N}\! \in \mathbb{C}^{n \times 1}=\{x_{1}, \ldots, x_{n}\}$ as:
\begin{equation}
\textbf{x}_1^N = \textbf{u}_1^N \cdot \textbf{F}^{ \otimes N}, \tag{1}
\end{equation}
where $\textbf{u}_1^N$ is the source bit, $\textbf{F}^{ \otimes N}$is the \textit{Kronecker} product of base matrix $\textbf{F} =\begin{bmatrix}1 & 0\\ 1 & 1\end{bmatrix}$ , and $\textbf{x}_1^N$ is the encoded bit.

\subsection{Belief Propagation Based Decoding of Polar Codes}
The BP decoding algorithm is based on messages passing between check nodes and variable nodes on both sides of the factor graph. Polar codes are decoded through iterative two-way message updates. \cite{Pamuk2011} proposed a hardware-friendly structure where messages are propagated and updated iteratively from the right to the left side as follows:
\begin{align}
L_{i,j}= f \left(L_{i+1,2j-1},g \left(L_{i+1,2j},R_{i,j+N/2} \right) \right), \tag{2}\label{eq2}\\
L_{i,j+N/2}= g \left(f \left(R_{i,j}, L_{i+1,2j-1} \right), L_{i+1,2j} \right), \tag{3}\label{eq3}\\
R_{i+1,2j-1}= f \left(R_{i,j}, g \left(L_{i+1,2j}, R_{i,j+N/2} \right) \right), \tag{4}\label{eq4}\\
R_{i+1,2j}= g \left(f \left(R_{i,j}, L_{i+1,2j-1} \right), R_{i,j+N/2} \right), \tag{5}\label{eq5}
\end{align}where $L_{i,j}$and $R_{i,j}$ are the messages propagated left or right of the node in row $i$ and column $j$, respectively. According to the min-sum algorithm, $f(x, y)\approx {\rm sign} (x) \cdot {\rm sign} (y) \cdot \min(\vert x\vert, \vert y\vert)$ and $g(x,\ y)=x+y$.

\subsection{Stochastis Computing}
\begin{figure}[tp]    
  \centering            
  \subfloat[]   
  {
      \label{fig1:subfiga}\includegraphics[width=0.24\textwidth]{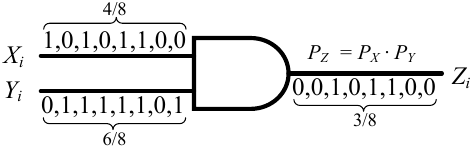}
  }
  \subfloat[]
  {
      \label{fig1:subfigb}\includegraphics[width=0.24\textwidth]{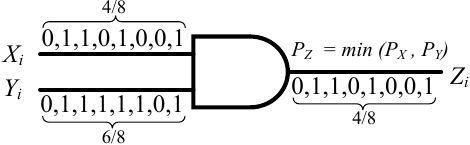}
  }
  \caption{AND gate function in different SCC. (a) SCC=0; (b) SCC=+1.}    
  \label{fig1}            
\end{figure}

In stochastic computing, a message is represented by the probability of a `1' occurring in a bit stream. The correlation between bit streams affects the calculation results, and the stochastic computation correlation (SCC) \cite{Alaghi2013} is defined by: 

\begin{align*}
\mathrm {SCC}(x,y) = \begin{cases} {}\frac {\delta (x,y)}{\min (p_{x},p_{y}) - p_{x}p_{y}}, & \delta (x,y) > 0 \\ 0, & \delta (x,y) = 0 \\ {}\frac {\delta (x,y)}{p_{x}p_{y} - \max (p_{x} + p_{y} - 1,0)}, & \delta (x,y) < 0 \\ \end{cases}\tag{6}
\end{align*}

\noindent
where $\delta (x,y)= p_{xy}-p_xp_y$, with $p_{xy}$ representing the probability of the output from an AND gate with inputs $x$ and $y$.  
Different SCC values represent specific relationships. $\mathrm {SCC} = 0$ means the bit streams are independent. $\mathrm {SCC} = +1$ means they are positively correlated, implying that they can be considered generated from the same random number sequence. And $\mathrm {SCC} = -1$ means the bit streams are negatively correlated. The function of the circuit varies with different SCC values, as illustrated in Fig. \ref{fig1}. For example, an AND gate functions as a multiplier when $\mathrm{SCC}=0$ and as a minimum value calculator when $\mathrm{SCC}=+1$. Other stochastic logic circuit functions under different relationships are shown in Table \ref{tab1}.

\begin{table}[t]
    \centering
    \renewcommand\arraystretch{1.3}
    \caption{Logic Circuit Functions Under Different Relationships}
    \begin{tabular}{c||c|c}
    \toprule[1pt]
        Relationship & Logic Circuit & Function \\ \hline\hline
                    & AND & \( z = x \cdot y \) \\ \cline{2-3}
        Independent & OR & \( z = (x + y) - x \cdot y \) \\ \cline{2-3}
                    & XOR & \( z = x \cdot (1 - y) + (1 - x) \cdot y \) \\ \hline
                    
                                  & AND & \( z = \min(x, y) \) \\ \cline{2-3}
        Positively Correlated     & OR & \( z = \max(x, y) \) \\ \cline{2-3}
                                  & XOR & \( z = |x - y| \) \\ \hline           
                                  
                                  & AND & \( z = \max(x - y, 0) \) \\ \cline{2-3}
       Negatively Correlated      & OR & \( z = \min(x + y, 1) \) \\ \cline{2-3}
                                  & XOR & \( z = \min(x+y,2-(x+y))\) \\ 
\bottomrule[1pt]
    \end{tabular}
    \label{tab1}
\end{table}

\section{Correlated Stochastic Decoder}
\label{sec:3}
In this section, we propose a CU based on correlated stochastic computing to facilitate the iteration of the min-sum algorithm. Additionally, we introduce several optimized designs tailored for the correlated stochastic decoder, aiming to enhance decoding performance, reduce the number of iterations, and minimize circuit area.
\subsection{Computing Unit}
\begin{figure}[t]
  \centering
  \subfloat[]
  {\includegraphics[width=0.23\textwidth]{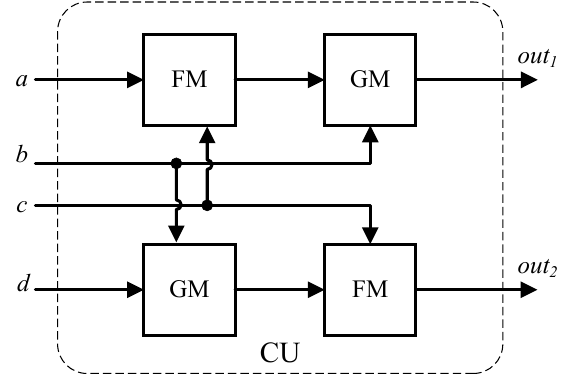}\label{fig2:subfiga}}
  \subfloat[]
  {\includegraphics[width=0.23\textwidth]{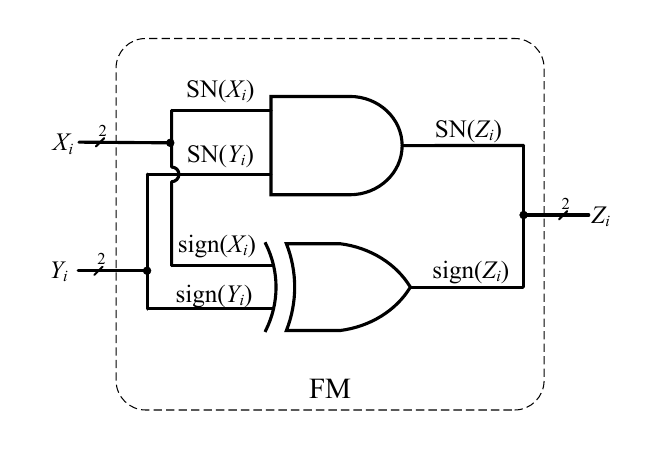}\label{fig2:subfigb}}\\
  \subfloat[]
  {\includegraphics[width=0.23\textwidth]{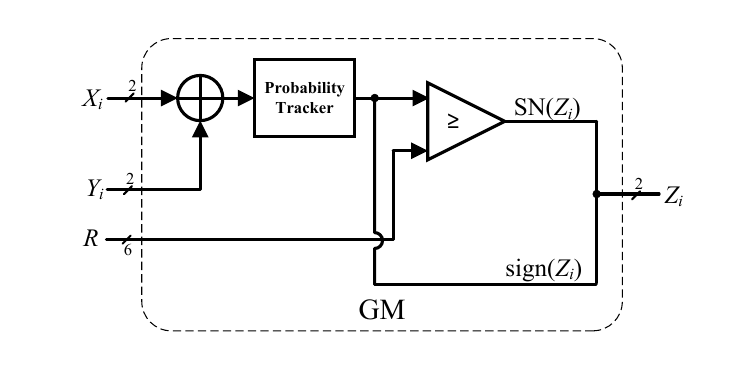}\label{fig2:subfigc}}
  \subfloat[]
  {\includegraphics[width=0.23\textwidth]{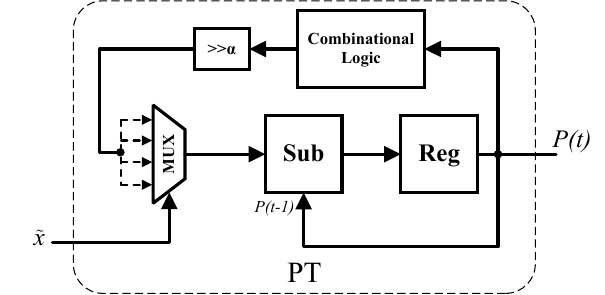}\label{fig2:subfigd}}\\
  \caption{The structure of (a) computing units; (b) $F$ function module; (c) $G$ function module; (d) probability tracker.}
\end{figure}

The decoding of polar codes is achieved through iterative processing of equations \eqref{eq2} - \eqref{eq5}. Previous work has shown that using a round-trip approach with alternating left-to-right iterations is a more effective update method\cite{Xu2015}. Therefore, we designed a unidirectional-output CU, which updates only one type of message at a time, as illustrated in Fig. \ref{fig2:subfiga}. 
FM and GM are stochastic implementations used to realize the functions of the min-sum algorithm. In Fig. \ref{fig2:subfigb}, the f-function requires only one AND gate and one XOR gate, significantly reducing area and cost. To implement the g-function, a probability tracker (PT) is employed, which couples the iterative and computational processes, thereby greatly reducing latency. The update equation of PT in the GM is: 
\begin{align}
P(t)= (1-\alpha)\cdot P(t-1)+\alpha\cdot \tilde{x}.\tag{7}\label{eq7}
\end{align}
To avoid the additional complement operations caused by the polarity of $x$, equation \eqref{eq7} can be rewritten as: 
\begin{align*}
P(t)= P(t-1)-\alpha\cdot[P(t-1)-\tilde{x}],\tag{8}\label{eq8}
\end{align*}
where $\alpha$ is the relaxation coefficient\cite{Sharifi2010}, expressed as $2^{-m}$ to eliminate the need for multipliers. Considering that the value of $\tilde{x}$ after passing through the adder and inputting into PT can be $\{-2,-1,0,1,2\}$, meiosis can be optimized by combinatorial logic simplification, calculated as equations \eqref{eq12}, where $P_{n}$ is the $n$-th bit of the probability value stored in the register and input to the combinatorial logic, $Q_{n}$ is the $n$-th bit of the combinatorial logic output, and $Q_{n-2}$ to $Q_{0}$ remain unchanged. This optimization reduces the area of an $n$-bit adder, when using $n$-bit quantization. After optimization, PT is shown in Fig. \ref{fig2:subfigd}, achieving a smaller area 9\%\cite{Sharifi2010}. 

\begin{align*}
Q_{n-1}& = \begin{cases}
    P_{n-1}                          & \tilde{x}=-2 \\
    \overline{P_{n-1}}                 & \tilde{x}=-1 \\
    \overline{P_{n-1} }                   & \tilde{x}=+1 \\
    P_{n-1}                          & \tilde{x}=+2 
\end{cases}\;\;\;\;
Q_{n} = \begin{cases}
    \overline{P_{n} }                       & \tilde{x}=-2 \\
    P_{n}\oplus P_{n-1}                & \tilde{x}=-1 \\
    \overline{P_{n}\oplus P_{n-1}}     & \tilde{x}=+1 \\
    \overline{P_{n}  }                      & \tilde{x}=+2 
\end{cases} 
\end{align*}

\begin{align*}
Q_{n+1}& = \begin{cases}
    0                             &  \tilde{x}=-2 \\
    P_{n}\cdot \overline{P_{n-1}} &  \tilde{x}=-1 \\
    P_{n}+\overline{P_{n-1}}      &  \tilde{x}=+1 \\
    1                             &  \tilde{x}=+2 
\end{cases} \tag{9}\label{eq12}
\end{align*}



\begin{table}[tp]
\caption{Reduction Ratios At Different Code Lengths}
\centering\centering
\renewcommand\arraystretch{1.3} 
\begin{tabular}{c||ccccc}
\toprule[1pt]
\multirow{2}{*}{\begin{tabular}[c]{@{}c@{}}Reduction\\ Ratio\end{tabular}} & \multicolumn{5}{c}{Code Length}                                                                                                          \\ \cline{2-6} 
                                 & \multicolumn{1}{c|}{$N=32$}   & \multicolumn{1}{c|}{$N=64$}   & \multicolumn{1}{c|}{$N=128$}  & \multicolumn{1}{c|}{$N=256$}  & $N=512$  \\ \hline\hline
FM                               & \multicolumn{1}{c|}{$32.5\%$} & \multicolumn{1}{c|}{$27.1\%$} & \multicolumn{1}{c|}{$27.1\%$} & \multicolumn{1}{c|}{$28.2\%$} & $28.1\%$ \\ \hline
GM                               & \multicolumn{1}{c|}{$13.1\%$} & \multicolumn{1}{c|}{$10.9\%$} & \multicolumn{1}{c|}{$12.2\%$} & \multicolumn{1}{c|}{$14.2\%$} & $14.8\%$ \\ 
\bottomrule[1pt]
\end{tabular}
\label{tab3}
\end{table}

\subsection{Frozen Factor Graph}
\begin{figure}[t]
  \centering
  {\includegraphics[width=0.48\textwidth]{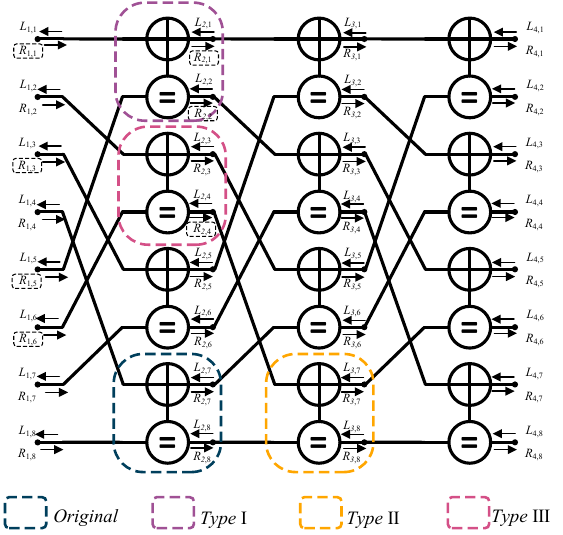}}
  \caption{Factor graph with $N=8$ and CU type simplified based on frozen bits. }
  \label{fig3}
\end{figure}

Polar codes enhance the reliability of each subchannel through polarization effects. As a result, message bits are carried on good channels, while frozen bits are allocated to poor channels based on the reliability ranking of the subchannels. Frozen bits do not require message updates, and this characteristic propagates through the factor graph, as illustrated in Fig. \ref{fig3}. Depending on the presence of frozen bits, we have designed three simplified CUs for L message updates and R message updates, respectively, to reduce circuit area and accelerate decoding iteration.

\begin{figure}[t]
  \centering
  \subfloat[]
  {\includegraphics[width=0.16\textwidth]{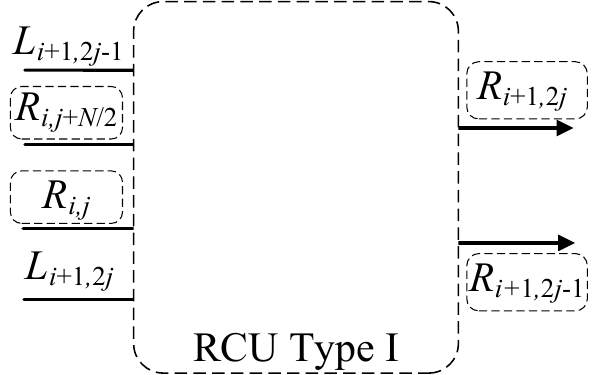}\label{fig4:subfiga}}
  \subfloat[]
  {\includegraphics[width=0.16\textwidth]{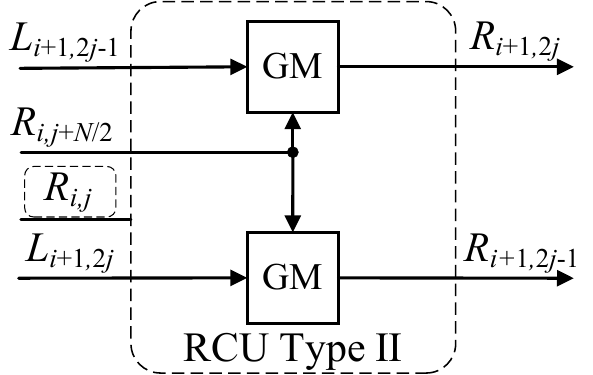}\label{fig4:subfigb}}
  \subfloat[]
  {\includegraphics[width=0.16\textwidth]{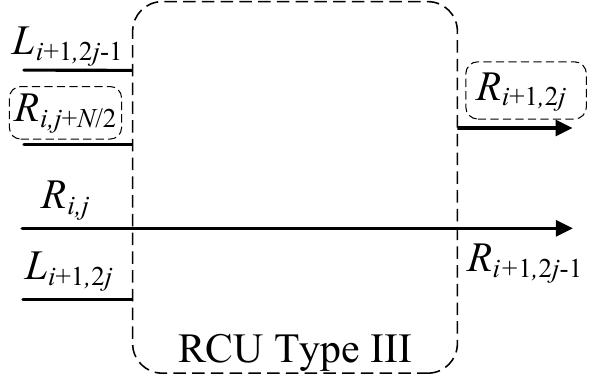}\label{fig4:subfigc}}\\
  \subfloat[]
  {\includegraphics[width=0.16\textwidth]{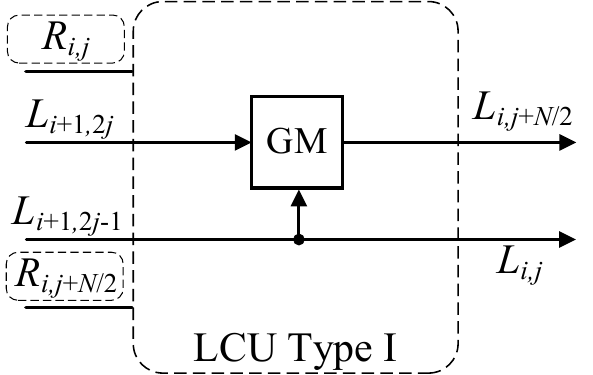}\label{fig4:subfigd}}
  \subfloat[]
  {\includegraphics[width=0.16\textwidth]{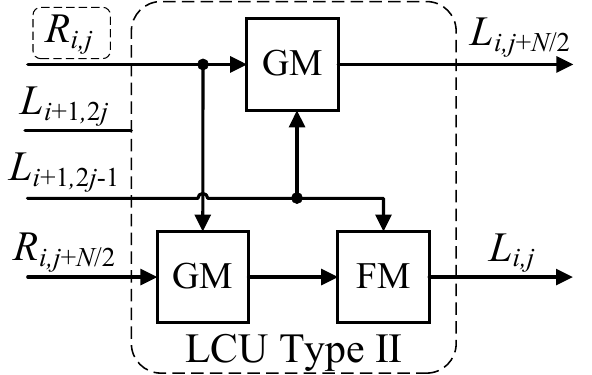}\label{fig4:subfige}}
  \subfloat[]
  {\includegraphics[width=0.16\textwidth]{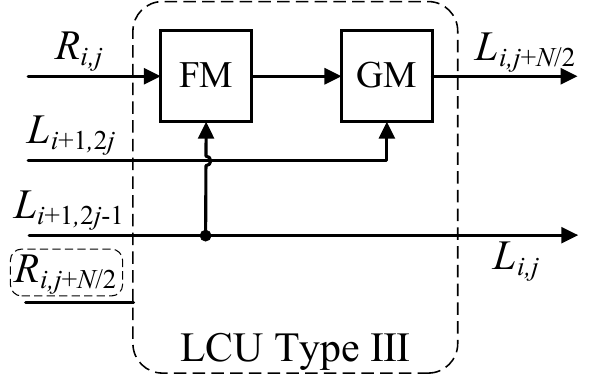}\label{fig4:subfigf}}\\
  \caption{The structure of simplified CUs. Type $\mathrm{I}$: both $R_{i,j}$ and $R_{i,j+N/2}$ are frozen bits; Type $\mathrm{II}$: $R_{i,j}$ is frozen bits; Type $\mathrm{III}$: $R_{i,j+N/2}$ is frozen bits. }\label{fig4}
\end{figure}

In Fig. \ref{fig3}, multiple CUs are identified based on different situations involving frozen bits. When updating R messages, both type 1 and type 3 CUs can propagate frozen messages, simplifying two FMs and two GMs and greatly reducing circuit complexity. Additionally, the R messages on the left side of the frozen bits is set to the maximum real value to ensure the propagation of prior information in the min-sum algorithm. However, due to the limited representation range of random bit streams, this characteristic cannot be guaranteed during stochastic computing. Our redesigned CU does not update frozen messages but instead combines them with the update equations $f (x, y)$ and $g (x, y)$, as shown in Fig. \ref{fig4}. For a polar code of (256, 128), the optimized design can reduce FM by 28.2\% and GM by 14.2\%, using the improved degrading-merge algorithm (Algorithm D in \cite{Tal2013}). The reduction rate for codes of other lengths is shown in Table \ref{tab3}. The optimized decoder can reduce the number of iterations by 25.2\% at $E_b/N_0=3$ dB, as shown in Table \ref{tab2}.

\section{Hardware Implementation And Performance Evaluation}
\label{sec:4}

\begin{figure*}[t]
  \centering
  {\includegraphics[width=0.85\textwidth]{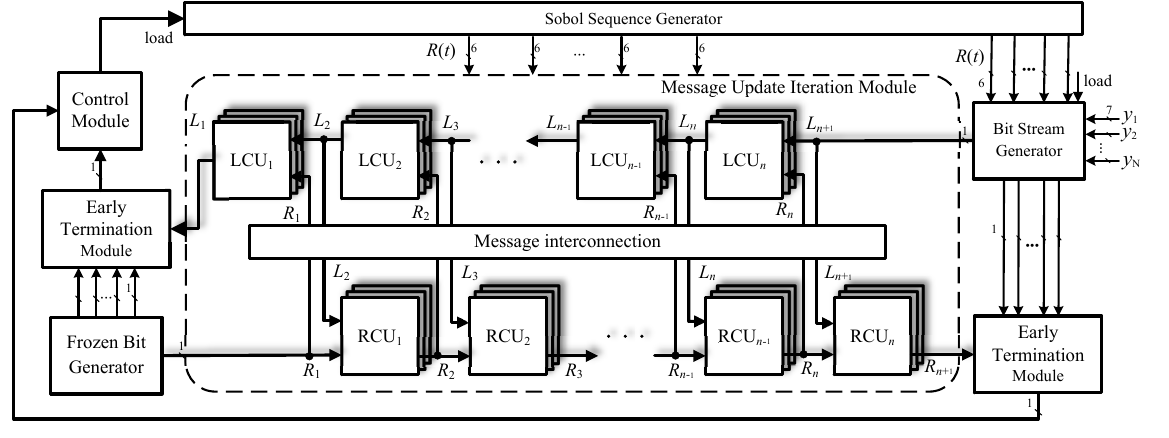}}
  \caption{The architecture of ECS-PD.}
  \label{fig5}
\end{figure*}

\subsection{The Structure Of ECS-PD}
Fig. \ref{fig5} illustrates the correlated stochastic polar decoder. 
The structure includes $n$ levels, each containing $N/2$ CUs, where $N$ is the length of the polar code and $n=\log_2 N$. Messages are input from both ends, consisting of $R_{1}$ bit stream generated based on frozen bits, generated based on the LLR received through the channel. The left propagation messages $L_{n+1}$ of the $(n+1)$-th column node and the right propagation messages $R_{1}$ of the $1$-th column node are represented by equations \eqref{eq10} and \eqref{eq11} respectively, where $R(t)$ represents the value of the random bit stream source.

Previous work focused on stream update solutions, which resulted in slow message update speeds and slow convergence of iterations. This paper introduces a PT into the decoder, integrating the iterative process and message update process. Consequently, messages no longer need to be represented by fixed-length random bit streams. In the architecture depicted in Fig. \ref{fig5}, bit streams flow separately in the LCU and RCU and are transmitted between them through the interconnection module. As a result, the proposed message update iteration module does not require block memory, such as RAM, and updates messages in a pipeline fashion.

In previous works\cite{Sharifi2006}\cite{Xu2020}, a scale factor was introduced to address the issue of the SCC approaching +1 or -1 in high SNR states, calculated in the probability domain. However, the ECS-PD operates in the LLR domain and does not encounter this problem. To prevent data saturation from excessive LLR or accuracy loss due to proportional scaling, this paper also introduces a scale factor $ N_0 $, where $N_0$ is the single-sided noise power density.

\begin{align*}
LLR_{j}^{\prime} &= N_0 \cdot LLR_{j}\\
&=  N_0 \cdot \log \left[\frac{\exp \left(-(y_{j}-1)^2/N_{0}\right)}{\exp \left(-(y_{j}+1)^2/N_{0}\right)}\right]\nonumber\\[-1pt]
&= 4\cdot y_{j} , \tag{10}\\\\
L_{n+1,j} &= \begin{cases}  
    1, & R(t)\leq\mid LLR_{j}^{\prime} \mid , LLR_{j}^{\prime}>0  \\ 
    -1, & R(t)<\mid LLR_{j}^{\prime} \mid , LLR_{j}^{\prime}<0  \\ 
    0, & R(t)>\mid LLR_{j}^{\prime} \mid  
\end{cases}\tag{11}\label{eq10} \\\\
R_{1,j} & = \begin{cases}  
    1, & u_j \in frozen\; bit  \\ 
    0, & u_j \notin frozen\; bit  
\end{cases}\tag{12}\label{eq11}
\end{align*}

\begin{table}[t]
\caption{Average Iteration Times Of (256,128) Polar Codes Implemented By Different CUs}
\centering\centering
\renewcommand\arraystretch{1.3} 
\begin{tabular}{c||cccccc}
\toprule[1pt]
\multirow{2}{*}{\begin{tabular}[c]{@{}c@{}}CU\\ Type\end{tabular}} & \multicolumn{6}{c}{Eb/N0}                                                                                                          \\ \cline{2-7} 
                                    & \multicolumn{1}{c|}{$2.5\,\mathrm{dB}$}   & \multicolumn{1}{c|}{$3\,\mathrm{dB}$}  & \multicolumn{1}{c|}{$3.5\,\mathrm{dB}$}  & \multicolumn{1}{c|}{$4\,\mathrm{dB}$}& \multicolumn{1}{c|}{$4.5\,\mathrm{dB}$}  & $5\,\mathrm{dB}$  \\ \hline\hline
Original                               & \multicolumn{1}{c|}{$360.62$} & \multicolumn{1}{c|}{$234.21$} & \multicolumn{1}{c|}{$122.82$} & \multicolumn{1}{c|}{$72.73$} & \multicolumn{1}{c|}{$37.39$} & $22.58$ \\ \hline
Simplified                               & \multicolumn{1}{c|}{$315.64$} & \multicolumn{1}{c|}{$175.12$} & \multicolumn{1}{c|}{$108.80$} & \multicolumn{1}{c|}{$55.70$} & \multicolumn{1}{c|}{$28.86$} & $17.74$ \\ 
\bottomrule[1pt]
\end{tabular}
\label{tab2}
\end{table}

\begin{figure*}[htbp]
  \centering
  \subfloat[]
  {\includegraphics[width=0.23\textwidth]{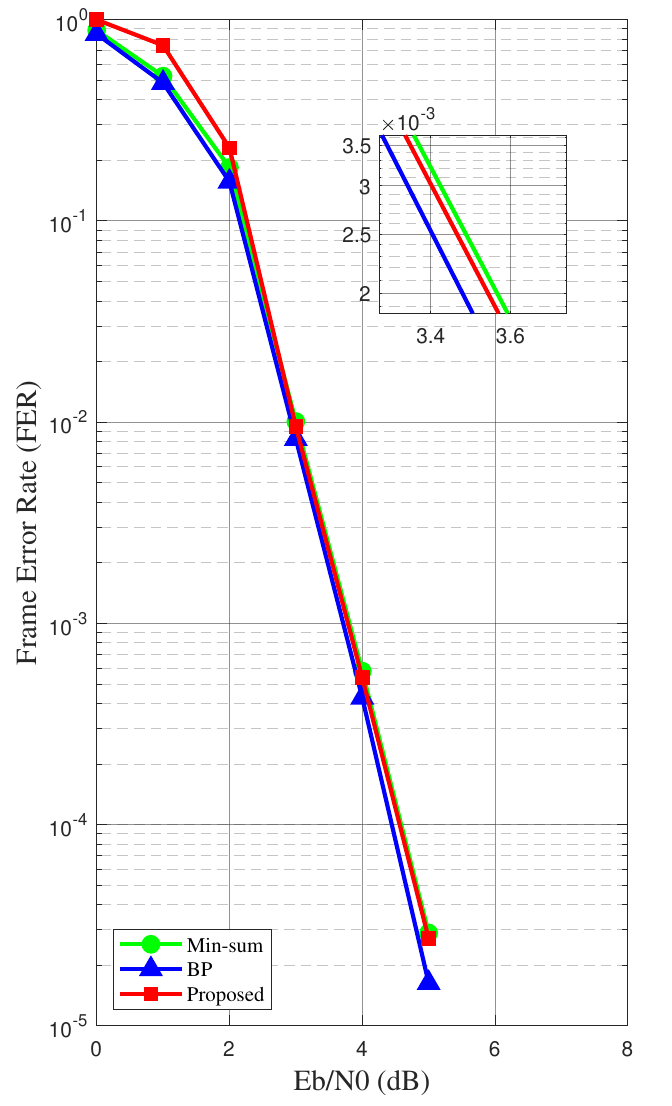}\label{fig7:subfiga}}
  \subfloat[]
  {\includegraphics[width=0.23\textwidth]{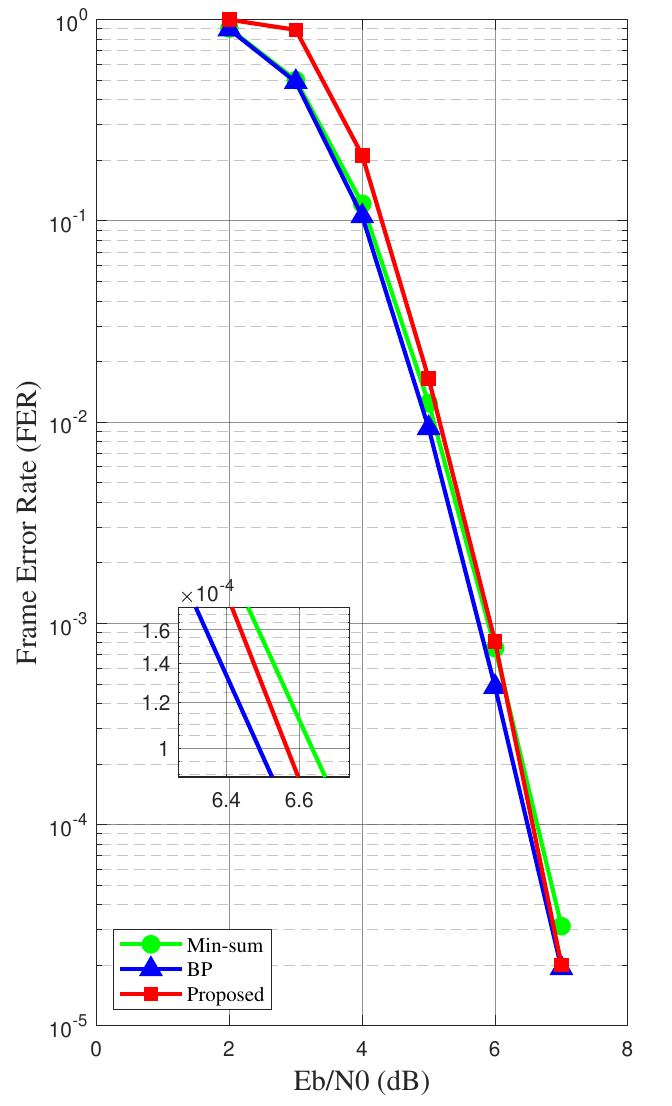}\label{fig7:subfigb}}
  \subfloat[]
  {\includegraphics[width=0.23\textwidth]{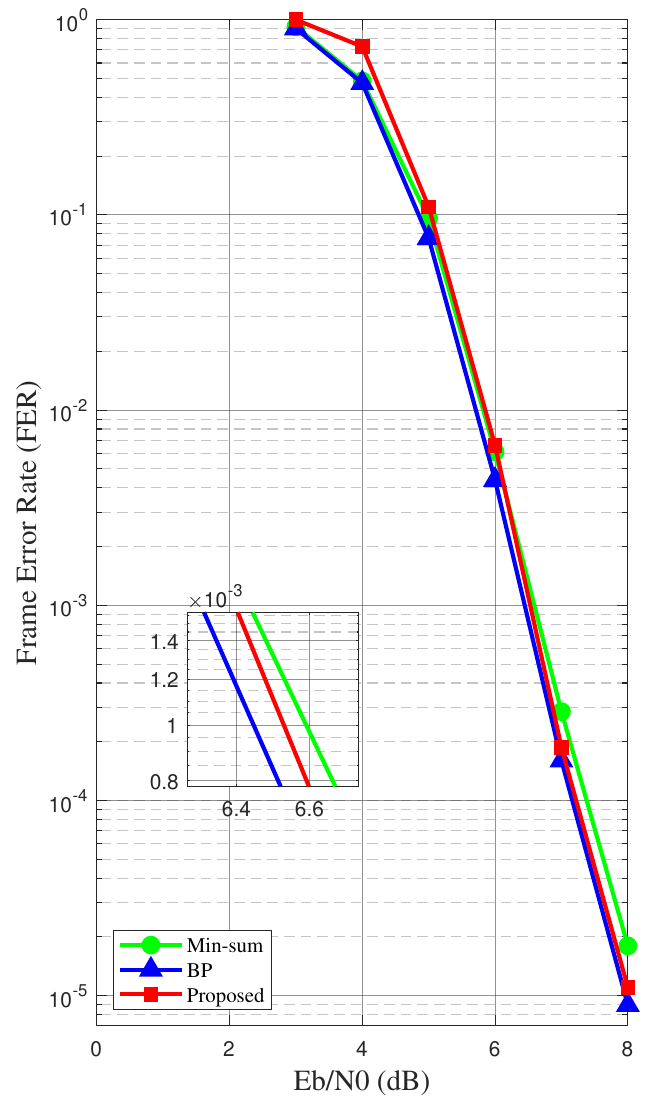}\label{fig7:subfigc}}
  \subfloat[]
  {\includegraphics[width=0.23\textwidth]{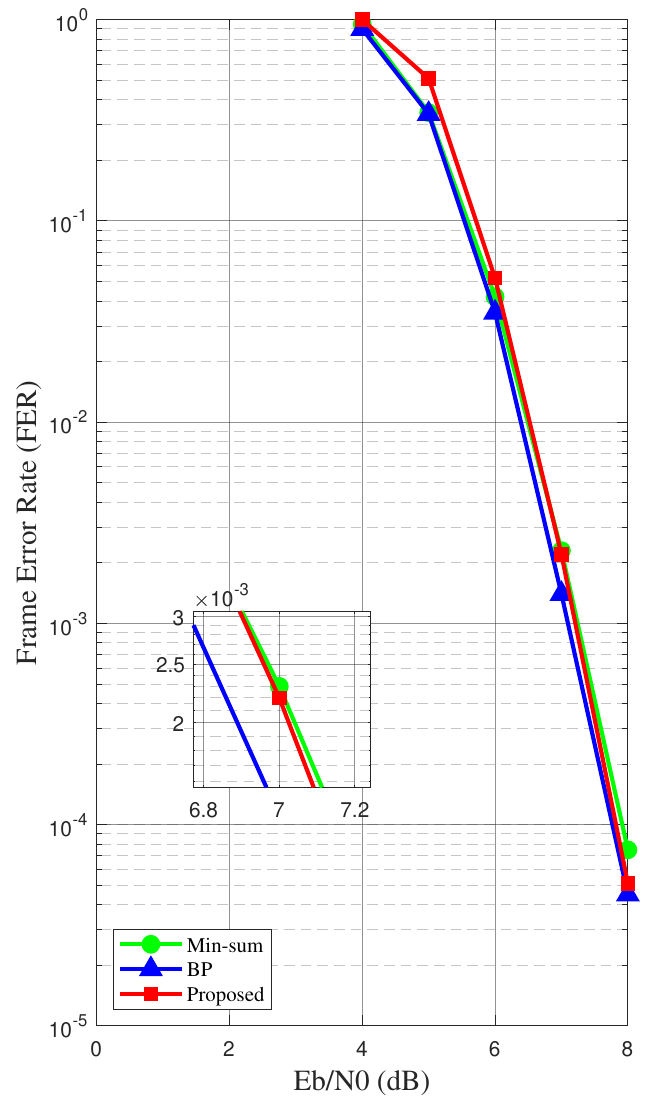}\label{fig7:subfigd}}\\
  \caption{The FER simulation result for different polar codes. (a) $N=256, K=128$, and code rate $R=\frac{1}{2}$; (b) $N=256, K=171$, and code rate $R\approx \frac{2}{3}$; (c) $N=256, K=192$, and code rate $R\approx \frac{3}{4}$; (d) $N=256, K=205$, and code rate $R\approx \frac{4}{5}$. }\label{fig7}
\end{figure*}

This design utilizes positive correlation stochastic computing, where all messages are represented by the same random bit stream, requiring only a single random number generator for the entire circuit. In contrast, the traditional stochastic BP decoder requires a random number generator at each stage to ensure sequence decorrelation. While a re-randomization module was introduced in previous work \cite{Xu2020}, it still required multiple Linear Feedback Shift Registers (LFSRs) to maintain the decorrelation of the random number source. 

If the PT in Fig. \ref{fig2:subfigd} changes slowly over the last T clock cycles and is considered stable, the influence of random bits prior to these $T$ cycles on the PT value is relatively insignificant \cite{Sharifi2010}. Consequently, the PT value can be regarded as dependent solely on the distribution of `1' bits within these $T$ clock cycles. Using a uniform distribution of `1' bits in the random bit stream over these $T$ clock cycles can significantly enhance the PT's accuracy. To generate such a random bit stream, a more uniform Sobol sequence \cite{Liu2018} can be employed, improving the PTr's performance. The Sobol sequence is a quasi-random sequence used to produce low-variance sequences, primarily to enhance the computational efficiency and accuracy of high-dimensional numerical integration and Monte Carlo simulations.
This design utilizes only a Sobol sequence generator to maintain sequence correlation and requires only $\log_2 l$ registers, where $l$ is the length of the random bit stream. Compared to decoders using independent sequences for calculations, this approach reduces the circuit area of the random number generator by a factor of $2\cdot N$. 

This paper implements a bidirectional judgment approach to accelerate convergence and facilitate the early termination of iterations. To achieve this, a cyclic redundancy check (CRC)-aided early termination scheme, as described in \cite{Ren2015}, is employed. The updated bit stream messages $L$ and $R$ are directed to different early termination modules for decision-making. The judgment message is derived through the operation of the $g(x, y)$ function. Once the CRC check is successful, the control module terminates the decoding process early and outputs the result. Our tests show that using CRC-16 for the polar code with $N = 256$ satisfies the performance requirements. For longer polar codes, longer check sequences are necessary to prevent false positives.

\subsection{Performance Evaluation}
In this subsection, the performance evaluation results of the ECS-PD are presented in Fig. \ref{fig7}. The polar code has a code length $N=256$ and the code rates $R$ are $\frac{1}{2}$, $\frac{2}{3}$, $\frac{3}{4}$ and $\frac{4}{5}$, respectively. The source bits are modulated using binary phase-shift keying (BPSK) and transmitted through additive white Gaussian noise (AWGN) channels. Since related works aim to achieve similar performance to BP and min-sum algorithms, for comparison, we also simulated and plotted the results of the min-sum decoder \cite{Abbas2017} and the floating-point BP decoder \cite{Arikan2008}. The maximum number of iterations was set to 40. The received LLR messages were quantized to 7 bits, and the probability tracker width in the G Module was set to 6 bits. Additionally, the parameter $\alpha$ was set to $2^{-2}$. As the SNR increases and the frame error rate (FER) gradually decreases at different code rates, the simulation results demonstrate that the ECS-PD achieves similar performance to the min-sum decoder when the FER is below 0.1, with only a minimal difference compared to the BP decoder. This indicates that the proposed decoder maintains comparable error correction capability to the conventional BP decoder, while requiring fewer resources. 
 Fig. \ref{fig7} illustrates the robustness of the ECS-PD across various code rates, suggesting that the decoder performs efficiently not only at lower code rates but also in more demanding scenarios where higher code rates are required.

\begin{table*}[htbp]
\centering
\begin{threeparttable}[b]
\caption{Comparison Of Hardware Implementation Of Different Polar Decoders}
\renewcommand\arraystretch{1.3} 
\begin{tabular}{c||>{\centering\arraybackslash}p{2.6cm}|>{\centering\arraybackslash}p{2.6cm}|>{\centering\arraybackslash}p{2.6cm}|>{\centering\arraybackslash}p{2.6cm}}
\toprule[1.5pt] 
\textbf{Design}                                         & \begin{tabular}[c]{@{}c@{}}ECS-PD\\ {[}this work{]}\end{tabular} & \begin{tabular}[c]{@{}c@{}}Stochastic Decoder\\ {\cite{Xu2020}}\end{tabular} & \begin{tabular}[c]{@{}c@{}}Stochastic Decoder\\ {\cite{Yuan2016}}\end{tabular} & \begin{tabular}[c]{@{}c@{}}Min-Sum\\ {\cite{Abbas2017}}\end{tabular}  \\ \hline\hline
Technology (nm)                                  & 28                                                                                      & 65                                                                    & 65                                                                     & 65                                                       \\ \hline
Code ($N$,$K$)                                   & (256,128)                                                                               & (256,128)                                                             & (256,128)                                                              & (256,128)                                                  \\ \hline
Bit Stream Length                               & -                                                                                      & 500                                                                   & 1024                                                                                                       & -                     \\ \hline
Clock Frequency (MHz)                            & 1625 (700\tnote{a}\;)                                                                                     & 700                                                                   & 670                                                                    & 334                                                 \\ \hline
Area (mm$^2$)                     &0.149                                                                                         & 0.129                                                                 & 0.169                                                                  & 0.818                                                      \\ \hline
Maximum Decoding Latency (cycles)               & 800                                                                                        & 5003\tnote{b}                                                                   & -                                                                     & 640                                                       \\ \hline
Avg.Decoding Latency (cycles@3.5 dB)             &  109                                                                                       & 688                                                                   & 15360                                                                 & 61                                                         \\ \hline
Avg.Throughput (Mb/s@3.5 dB)                    & 3816.5 (1644\tnote{a}\;)                                                                                         & 297.6                                                                 & 11.6                                                                   & 1401.7                                                     \\ \hline
Hardware Eficiency (Mb/s/mm$^2$) & 25614 (4753\tnote{a}\;)                                                                                         & 2019                                                                  & 138                                                                    & 1713      \\                                            
\bottomrule[1.5pt]
\end{tabular}
\label{tab4}
\begin{tablenotes}
       \item [a] The normalization to the 65 nm technology node is based on the following formula: frequency $\propto s$, where $s = \frac{\text{technology}}{65 \text{nm}} $.
       \item [b] The maximum decoding latency is calculated based on the average decoding latency and the maximum number of iterations of 40. 
     \end{tablenotes}
\end{threeparttable}
\end{table*}

\subsection{Hardware Implementation}
To demonstrate the superiority of the proposed hardware, the architecture was implemented using Verilog HDL on TSMC 28-nm CMOS technology. We used Synopsys Design Compiler for synthesis and compared the results with some advanced decoders, as shown in Table \ref{tab4}. Due to the simplicity of the stochastic computing circuit and the flexibility of the decoding architecture, this design can operate at a 700 MHz clock frequency. The coupling of message updates, the iterative convergence process, and the simplification of the factor graph significantly improve the average decoding delay, resulting in a remarkable increase in throughput. Compared to the decoder presented in \cite{Xu2020}, the proposed design achieves a 5.5-fold increase in throughput and a 2.7-fold improvement in hardware efficiency. It also offers faster decoding efficiency with comparable performance to the min-sum decoder \cite{Abbas2017}.

\vspace{20pt}
\section{Conclusion}
\label{sec:5}
This paper presents a novel design for stochastic polar decoders, utilizing correlated random sequences and applying the min-sum algorithm in the LLR domain. The correlated stochastic computing approach fundamentally prevents the occurrence of a `hold-state'. Additionally, we redesigned the computation unit and proposed two methods to enhance decoding performance. The use of the Round-Trip iteration scheme and a unidirectional-output structure results in higher throughput and circuit efficiency while maintaining performance similar to that of the min-sum decoder. In future work, we will continue to explore additional optimizations to further enhance decoding performance.

\bibliography{main_bib}
\bibliographystyle{IEEEtran}

\end{document}